\documentclass[epj,twocolumn]{webofc}
\usepackage[varg]{txfonts}  
\woctitle{MPGD2015}

\begin{document}
\title{Study of Negative-Ion TPC Using $\mu$-PIC for Directional Dark Matter Search}

\author{Tomonori Ikeda\inst{1}\fnsep\thanks{\email{tomonori@stu.kobe-u.ac.jp}} \and
        Kentaro Miuch\inst{1} \and
        Atsuhiko Ochi\inst{1} \and
        Ryota Yakabe\inst{1} \and
        Takashi Hashimoto\inst{1} \and
        Ryosuke Taishaku\inst{1} \and
        ~Daniel P. Snowden-Ifft \inst{2} \and
        Jean-Luc Gauvreau\inst{2} \and
        Toru Tanimori\inst{3} \and
        Atsushi Takada\inst{3} \and
        Kiseki Nakamura\inst{3} 
        % etc.
}
\institute{Department of Physics, Kobe University, Rokkodaicho, Nada Hyogo 657-8501
\and
           Department of Physics, Occidental College, Los Angeles, CA 90041, USA \and
           Department of Physics, Kyoto University, Kitashirakawaoiwakecho, Sakyo, Kyoto 606-8502, Japan
          }

\abstract{%
Negative-ion time projection chambers(TPCs) have been studied for low-rate and high-resolution applications such as dark matter search experiments. Recently, a full volume fiducialization in a self-triggering TPC was realized. This innovative technology demonstrated a significant reduction in the background with MWPC-TPCs. We studied negative-ion TPC using the $\mu$-PIC+GEM system and obtained sufficient gas gain with CS$_{2}$ gas and SF$_{6}$ gas at low pressures. We expect an improvement in detector sensitivity and angular resolution with better electronics.

}

\maketitle

%%%%%%%%%%%%%%%%%%%%    Introduction  %%%%%%%%%%%%%%%%%%%%%%%%
\section{Introduction}
\ \ \ Weakly interacting massive particles(WIMPs) are one of the candidates for dark matter. Among the various direct search experiments, the detection of nuclear recoil tracks is thought to provide strong evidence of dark matter detection. Gas detectors are among the possible nuclear track detection methods, and several groups have developed detectors and performed underground direction-sensitive dark matter searches\cite{Tanimori,DRIFT,DMTPC,MIMAC,Nakamura}.

Since the expected event rate is extremely low, a dark matter detector needs to be large(cubic-meter scale). Therefore, the electron diffusions during drift are large compared with the track lengths of the recoil nuclei(up to several mm) in conventional TPCs. Since spatial and angular resolutions are deteriorated by diffusion, the drift length is limited to several tens of cm with electron-drift detectors\cite{DRIFT}. Transverse diffusion could be suppressed using a strong magnetic field parallel to the drift field but it would increase the cost and complexity of the system.

Negative-ion TPCs have been studied for low-rate and high-resolution applications\cite{NITPC}. The advantage of negative-ion TPCs is that the ion diffusions can be strongly suppressed at low pressures and without a magnetic field. The DRIFT (Directional Recoil Identification from Tracks) group has pioneered the study of negative-ion TPCs and demonstrated a strong potential for fine tracking performance\cite{DRIFT_2}. Recently a discovery of \textquotedblleft minority carriers\textquotedblright \ in CS$_{2}$ gas broadened its potential, and the measurement of absolute Z-position in self-triggering TPCs became possible. Minority carriers appeared after adding a few percent O$_{2}$ to the original gas. Although the mechanism is not completely understood, several types of negative ions are thought to drift each with slightly different velocities. Using this technique, DRIFT was able to significantly improve their detector sensitivity\cite{Battat}.

Electron capture, ion transport and electron release in the gas amplification process need to occur in  negative ion TPC. While  forming negative ions has been well studied theoretically and experimentally, the detachment process is less well understood\cite{NI}. Negative ion gases that give sufficient gas gains are very limited; carbon disulfide(CS$_{2}$) and nitromethane(CH$_{3}$NO$_{2}$) are one of the few possible choices\cite{nitro}. These gases are not very safe and are not easy to use in an underground laboratory.

A safe negative-ion gas has long been sought as a TPC gas, and recently a GEM-TPC with SF$_{6}$ negative ion drift gas was reported\cite{Nguyen}. In addition, they observed minority carriers without any additional gas. Unlike CS$_{2}$, SF$_{6}$ is non-toxic, and can be handled easily while retaining the same advantages as CS$_{2}$ gas. Fluorine has a large cross section for WIMP-nuclear spin-dependent interaction; therefore, it has good properties as a target. SF$_{6}$ is generally used as high voltage insulating gas. Its electron affinity is very large, approximately 1.1 eV. Therefore, we need strong electric fields for electron-detachment and gas multiplication to occur. Micro-pattern gaseous detectors(MPGDs) are known to have strong electric fields and would be a suitable device to use with SF$_{6}$ as a TPC gas.

NEWAGE (NEw generation WIMP-search with Advanced Gaseous tracking device Experiment) has been using a $\mu$-PIC based TPC for direction-sensitive dark matter search with CF$_{4}$ gas. We achieved a 90\% confidential level direction-sensitive spin-dependent cross section limit of 557 pb for a WIMP mass of 200 GeV/c$^{2}$\cite{Nakamura}. The main background restricting the detector sensitivity is radioactive contamination in the $\mu$-PIC. We initiated a new study to reduce the radioactive background using two approaches. The first approach is to develop a low background $\mu$-PIC. Glass fiber clothes used to reinforce the polyimide were found to contain certain amounts of  U and Th which radiate $\alpha$ particles. We are developing a $\mu$-PIC with a low background polyimide. The second approach is Z-fiducialization using negative ion gas. This is expected to effectively reduce the background because backgrounds from $\mu$-PIC should be localized at low Z positions. Increased angular resolution is also expected with a negative-ion gas. The negative-ion TPC would be a  good technology to construct large-volume ($\rm > 100\     m^{3}$) detectors in the future.

In this study, the first study of the $\mu$-PIC+GEM system with the negative ion gases CS$_{2}$ and SF$_{6}$ is reported. 

%%%%%%%%%%%%%%%%%%%   NEWAGE0.1C detector %%%%%%%%%%%%%%%%%%%%%
\section{NEWAGE0.1c Detector}
\begin{figure}
\centering
\sidecaption
\includegraphics[width=7cm,clip]{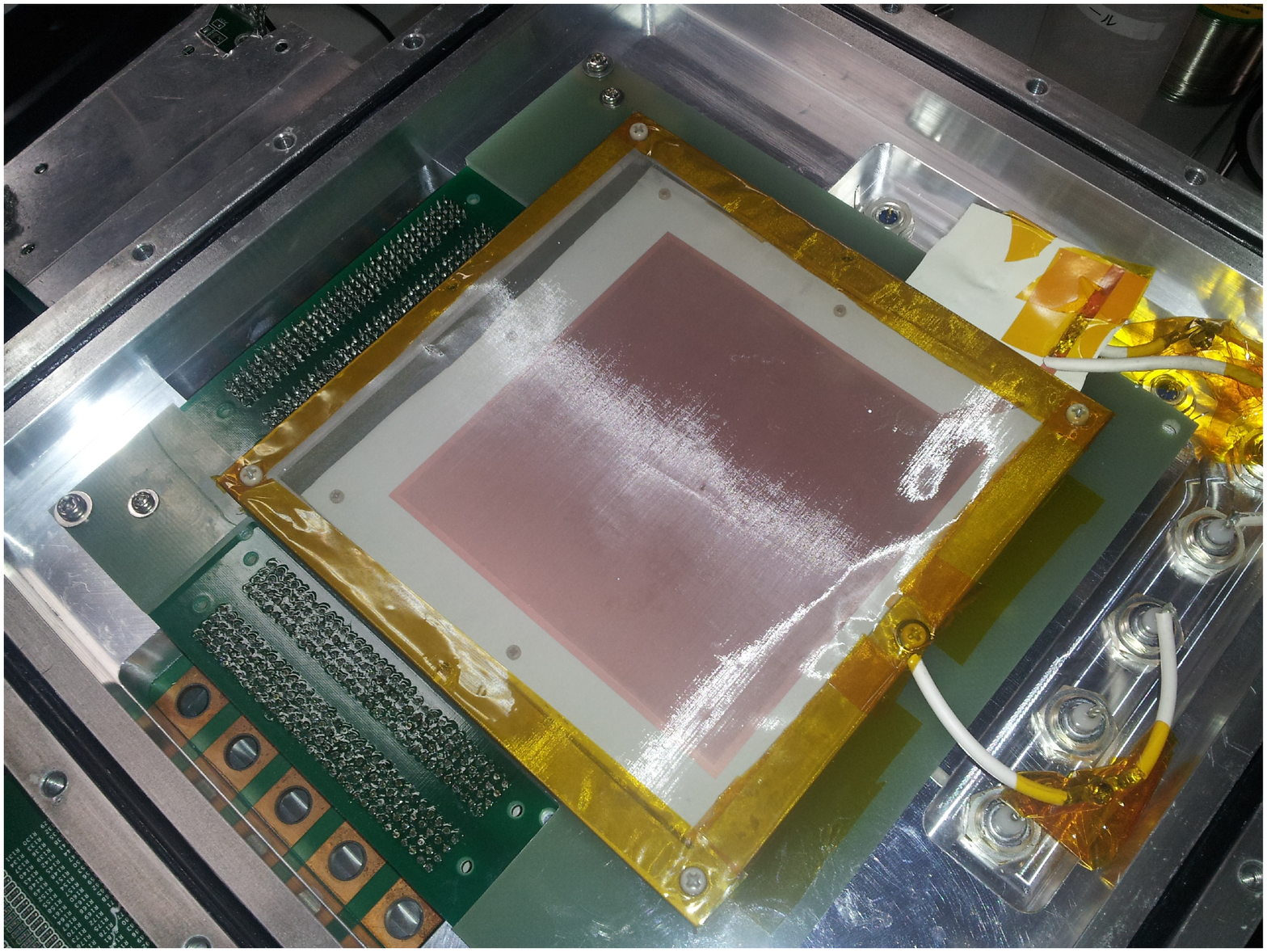}
\caption{Photograph of the NEWAGE0.1c detector. The upper side of the GEM electrode is seen through the drift mesh. The $\mu$-PIC is set below the GEM.}
\label{detector_figure}       % Give a unique label
\end{figure}

\begin{figure}
\centering
\sidecaption
\includegraphics[width=7cm,clip]{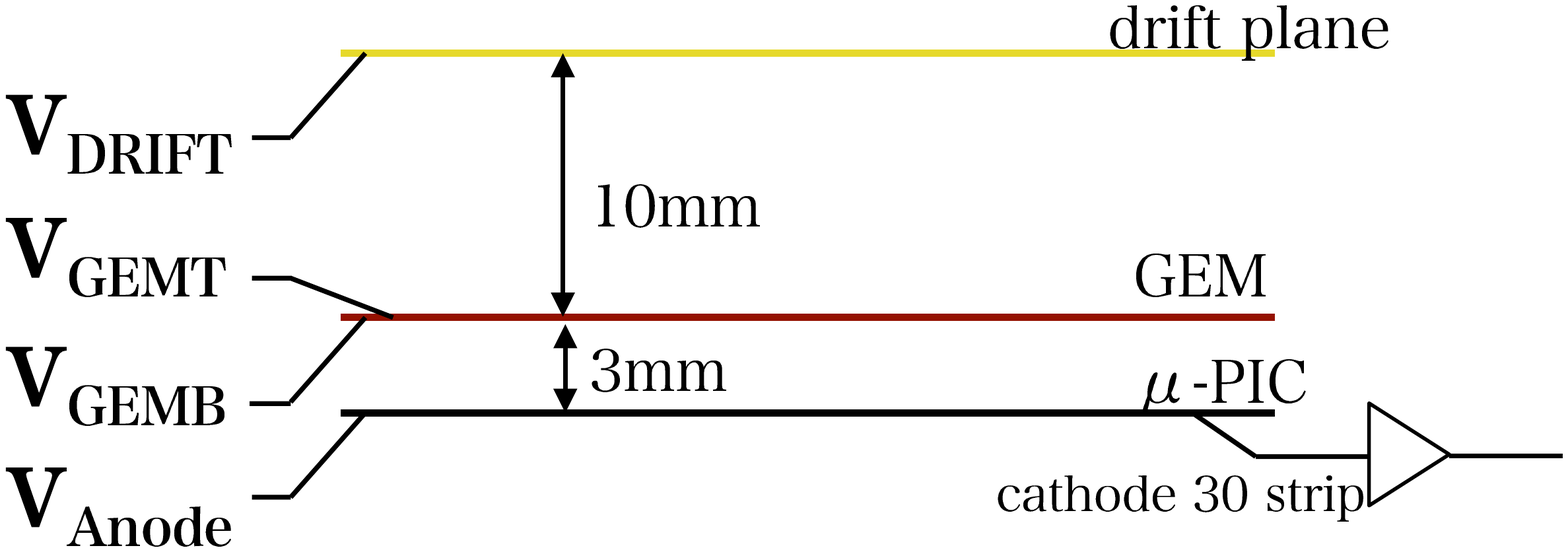}
\caption{Schematic structure of the NEWAGE0.1c. Some cathode strips are grouped and read by one channel amplifiers(see Section 3). V$_{\mathbf{DRIFT}}$ is the voltage supplied at the drift plane, V$_{\mathbf{GEMT}}$ is the voltage at the top of the GEM, V$_{\mathbf{GEMB}}$ is the voltage at the bottom of the GEM, and V$_{\mathbf{Anode}}$ is the voltage at the anode of the $\mu$-PIC. }
\label{detector}       % Give a unique label
\end{figure}

\ \ \ The NEWAGE0.1c detector which comprises of the $\mu$-PIC and the GEM(Fig.\ref{detector_figure}), was used for this work. A schematic view is shown in Fig.\ref{detector}. The $\mu$-PIC has 256 $\times$ 256 pixels with a pitch of 400 $\mu$m read by 256 anode and 256 cathode orthogonally formed strips\cite{u-PIC}.  When gas multiplication occurs at an anode, the same amount of charge is read from the corresponding anode and cathode strips. For this conceptual study, the sum of a few cathode strips was used as readout. The GEM\cite{GEM} was settled 3mm above the $\mu$-PIC as a sub-amplifier with an effective area of 10 $\times$ 10 cm$^{2}$. The substrate of the GEM was a 100 $\mu$m thick liquid crystal polymer(LCP) and the hole size and pitch were 70 $\mu$m and 140 $\mu$m respectively. A stainless steel mesh was set 10mm above the GEM as a drift plane. The $\mu$-PIC and the GEM were enclosed in a vacuum vessel. 

%%%%%%%%%%%%%%%%%%%  Measurement result  %%%%%%%%%%%%%%%%%%%%%%
\section{Measurement Results and Discussion}
\subsection{CS$_{2}$ gas}
\ \ \ We tested the $\mu$-PIC+GEM system with CS$_{2}$ gas using $^{55}$Fe at Occidental College. CS$_{2}$ gas is toxic, volatile, and inflammable; therefore, we performed the study in a dedicated gas-handling system. We supplied -1.5 kV at the drift plane. As the typical drift velocity of negative ions is about 10$^{-2}$ cm/$\mu$s, which is two or three orders of magnitude slower than electrons, we used a slow shaping amplifier(CREMAT,CR-111). The time constant was 4 $\mu$s and the gain was 130 mV/pC. The gas pressures were 76 Torr and 38 Torr. According to SRIM simulation\cite{SRIM}, the track length of a carbon nucleus with an energy of 100 keV is approximately 2 mm in CS$_{2}$ gas at 76 Torr which is longer than that in CF$_{4}$ gas at the same pressure.

\begin{figure}
\centering
\sidecaption
\includegraphics[width=7cm,clip]{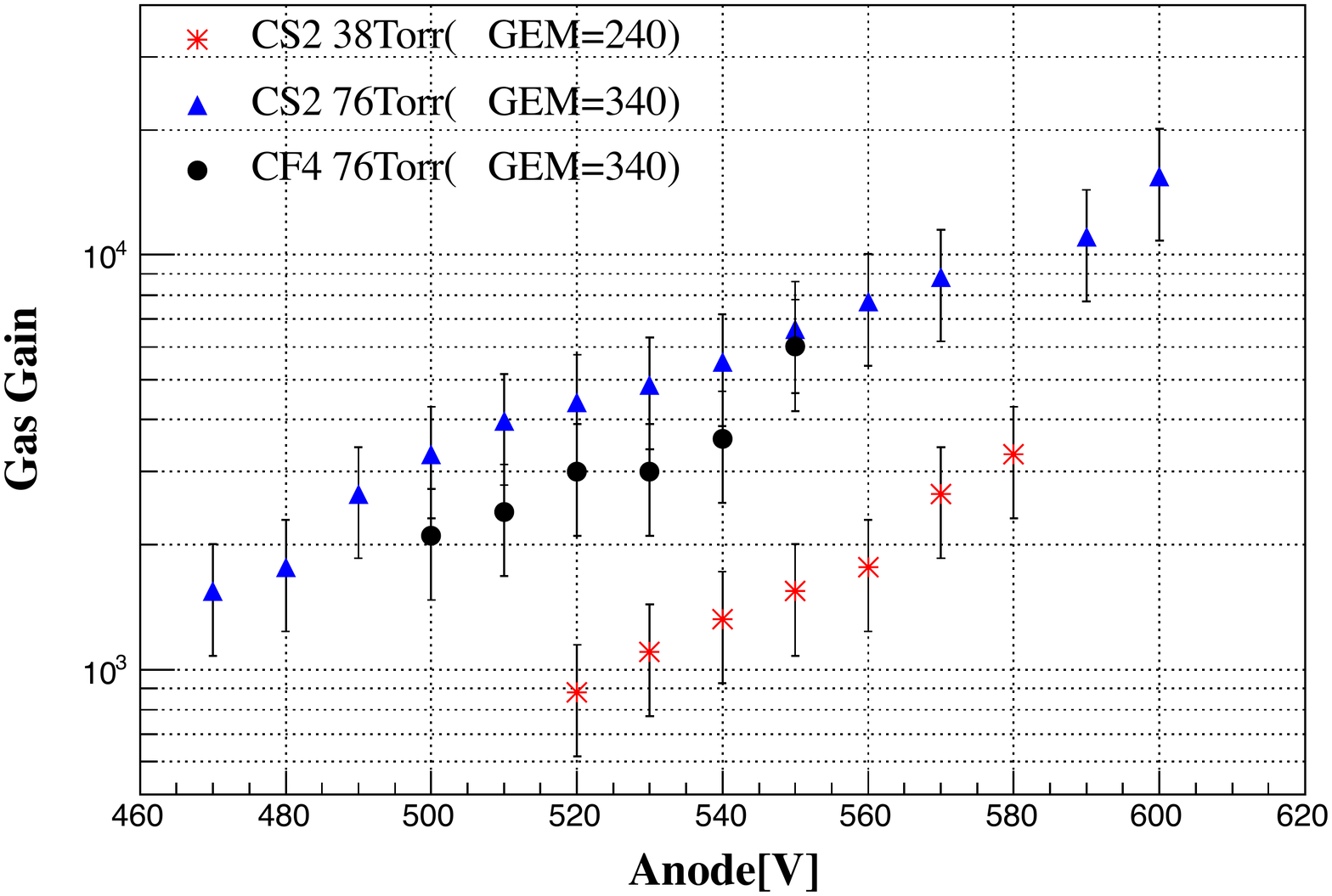}
\caption{Total gas gains as a function of the voltage supplied at the anode electrodes of the $\mu$-PIC. Circles, triangles and crosses show CF$_{4}$ at 76 Torr, CS$_{2}$ at 76 Torr and CS$_{2}$ at 38 Torr,  respectively.}
\label{CS2}       % Give a unique label
\end{figure}

\begin{figure}
\centering
\sidecaption
\includegraphics[width=7cm,clip]{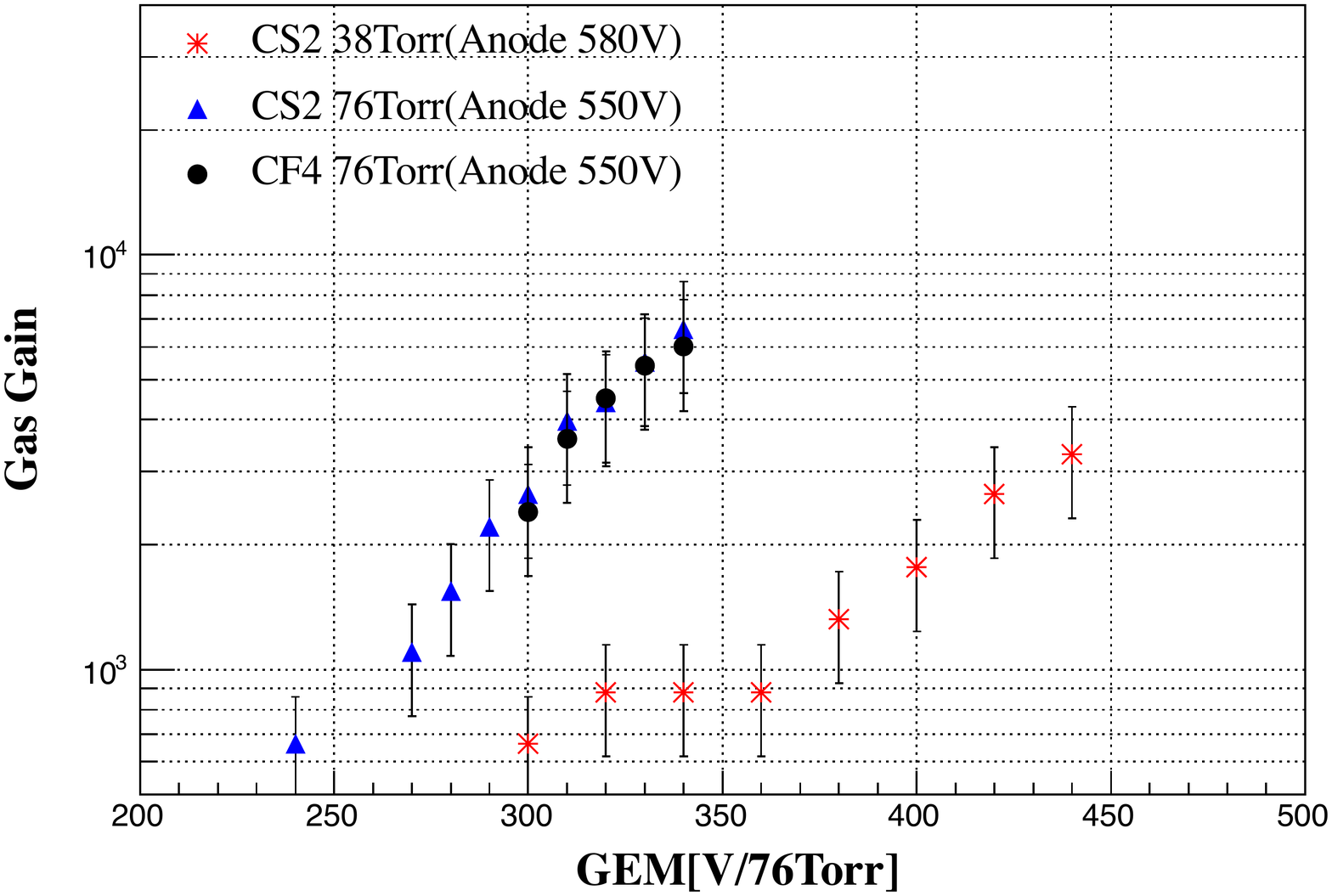}
\caption{Total gas gains as a function of the voltage supplied between the top and bottom electrodes of the GEM. Circles, triangles and crosses show CF$_{4}$ at 76 Torr, CS$_{2}$ at 76 Torr and CS$_{2}$ at 38 Torr, respectively.}
\label{CS2_2}       % Give a unique label
\end{figure}

The obtained total gas gain as a function of the voltage supplied at the anode electrodes of the $\mu$-PIC is shown in Fig.\ref{CS2}. The total gas gain of 10,000 at 76 Torr , was better than that achieved in CF$_{4}$ at the same pressure. Therefore, $\mu$TPC will have at least the same performance as CF$_{4}$ in low rate experiments. Total gas gains as a function of the voltage supplied to the GEM are shown in Fig.\ref{CS2_2}. We confirmed that the GEM performs as a sub-amplifier at both 76 Torr and 38 Torr. This total gas gain is sufficient to reach the energy threshold we recorded in previous experiments. 

Consequently, the $\mu$-PIC+GEM system worked very well with CS$_{2}$. We should be able to observe the minority carriers by adding O$_{2}$ gas.  The most remarkable point was achieving gas gains of a few thousand at 38 Torr. At this operating pressure, we can expect better angular resolution because the nuclear track is twice as long as with CF$_{4}$ at 76 Torr and track diffusion should be suppressed.  We expect that $\mu$TPC with CS$_{2}$ will have high angular resolution and high background rejection capabilities.

 %%%%% SF6 %%%%%
\subsection{SF$_{6}$ gas}
\ \ \ We tested the $\mu$-PIC+GEM system with SF$_{6}$ gas at a pressure of 20 Torr using $^{55}$Fe at Kobe University. We supplied -1.8 kV at the drift plane. We used a different slow shaping amplifier(CLEAR-PULSE,CS515-1)  for this experiment. The gain of CS515-1 is 1 V/pC and we coupled it with a high-pass filter that has $\tau_{decay}$ of $\sim$100 $\mu$s.

\begin{figure}
\centering
\sidecaption
\includegraphics[width=7cm,clip]{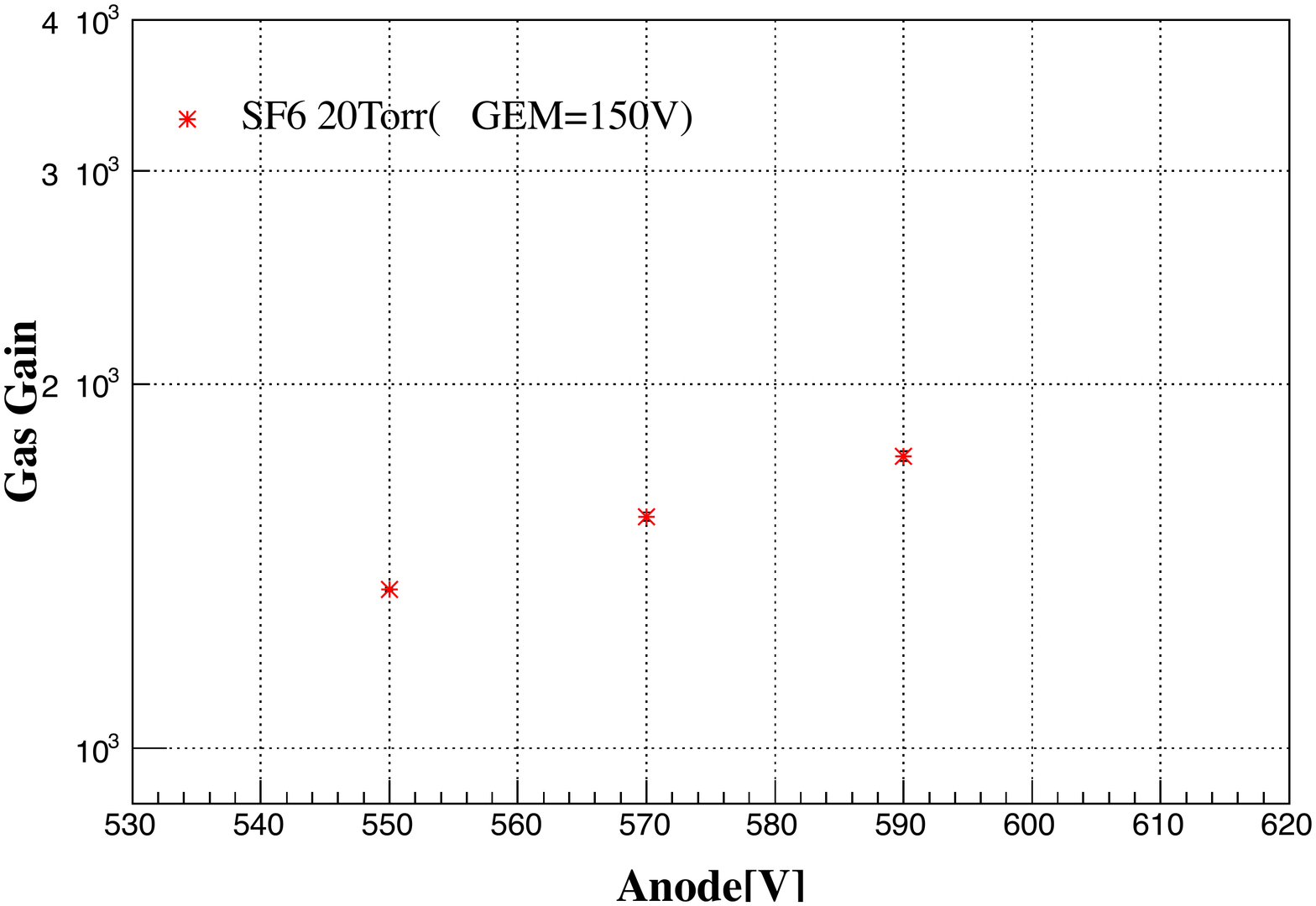}
\caption{Total gas gains as a function of the voltage supplied at the anode electrodes of the $\mu$-PIC. The voltages supplied at the top and bottom GEM are 840 V and 690 V, respectively. }
\label{SF6_anode_gain}       % Give a unique label
\end{figure}

\begin{figure}
\centering
\sidecaption
\includegraphics[width=7cm,clip]{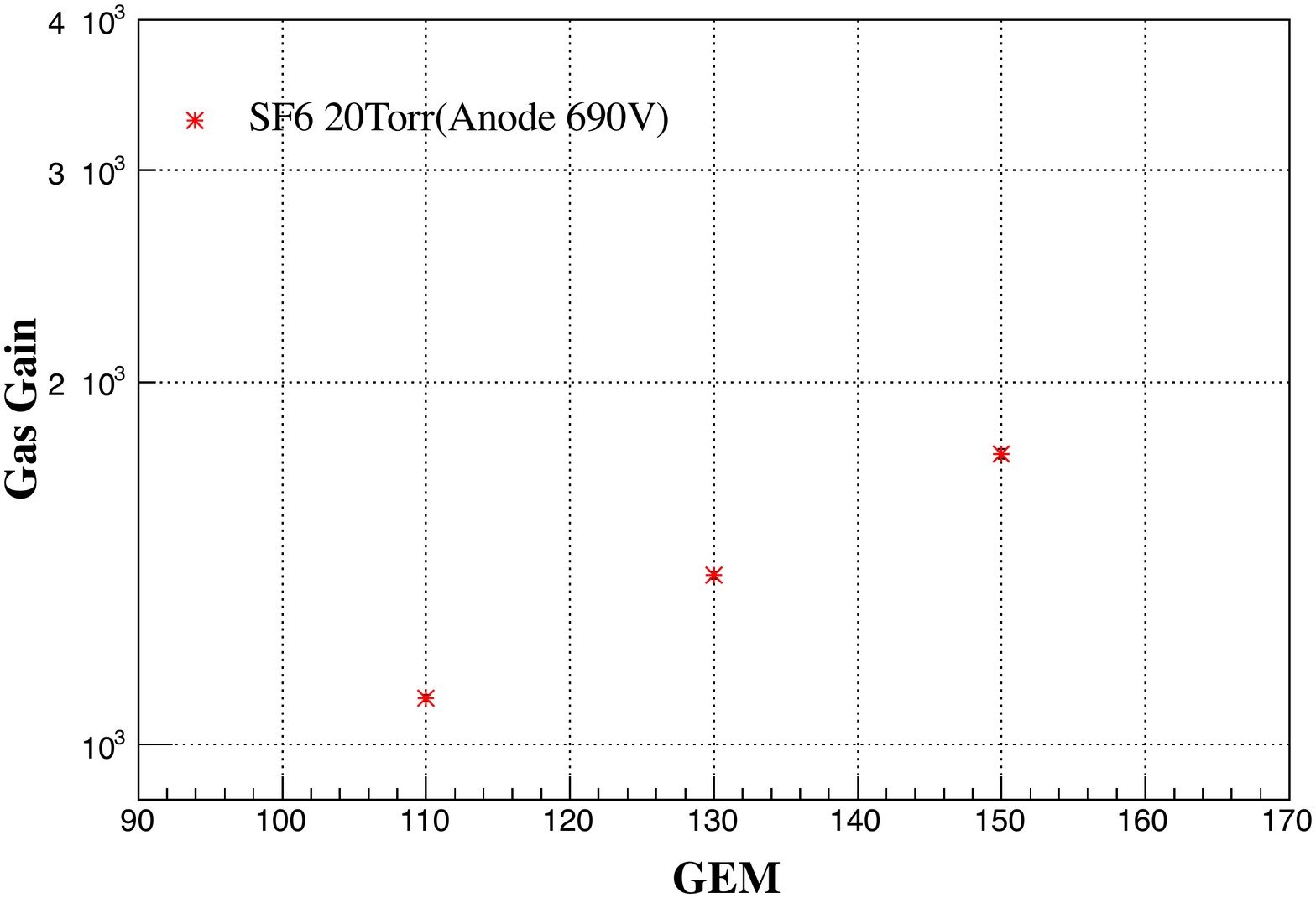}
\caption{Total gas gains as a function of the voltage supplied between the top and bottom electrodes of the GEM. The voltage supplied at the anode is 690 V.}
\label{SF6_GEM_gain}       % Give a unique label
\end{figure}

The obtained total gas gain as a function of the voltage supplied at the anode electrodes of the $\mu$-PIC and the voltage supplied to the GEM are shown in Fig.\ref{SF6_anode_gain} and Fig.\ref{SF6_GEM_gain}, respectively. The highest gas gain was approximately 1800, while the require gas gain was 1000 $\cdot$ $\frac{76\ Torr}{P}$. This means we need to improve the S/N by factor two, which could be achieved with a dedicated amplifier.

Consequently, the $\mu$-PIC+GEM system also worked well with SF$_{6}$. As noted in CS$_{2}$ experiments, we can expect better angular resolution because of  the longer nuclear track.

%%%%%%%%%%%%%%%%%%% Future work  %%%%%%%%%%%%%%%%%%%%%%
\section{Future work}
\ \ \ We have confirmed that the $\mu$-PIC+GEM system worked well with SF$_{6}$ and CS$_{2}$. For CS$_{2}$, we obtained  sufficient performance; however, because of gas safety issues, we will pursue studies of negative ion $\mu$TPC with SF$_{6}$. 

As our next step, we are constructing a prototype $\mu$TPC the drift length of which is 30 cm. We will attempt to observe the minority peak and to demonstrate measurement of the absolute Z position. Furthermore, we plan to develop readout electronics for negative ion $\mu$TPC. An ASIC chip with a multi-channel slow shaping amplifier is being developed for liquid argon TPC by the KEK e-sys group. This system can be adopted to negative ion TPCs with a small modification. We will then test the detector performance. We expect better spatial and angular resolution because of the smaller diffusion. A spatial resolution of $\sim$0.2 cm along the Z axis is demonstrated by the DRIFT detector\cite{Battat}; therefore we expect to reduce the background from the $\mu$-PIC by two orders of magnitude by rejecting events from the $\mu$-PIC plane. We expect an improvement of at least two orders of magnitude in the sensitivity because of full fiducialization.

%%%%%%%%%%%%%%%%%%% Acknowledgment  %%%%%%%%%%%%%%%%%%%%%%
\section{Acknowledgment}
This work was supported by KAKENHI Grant-in-Aids(19684005, 23684014 and 26104005) and Program for Advancing Strategic International Networks to Accelerate the Circulation of Talented Researches, JSPS, Japan(R2607).
%%%%%%%%%%%%%%%%%%%  Reference Paper  %%%%%%%%%%%%%%%%%%%%%%


\begin{thebibliography}{}
\bibitem{Tanimori} T. Tanimori et al., Phys. Lett. B \textbf{578} (2004)241-246
\bibitem{DRIFT} D.P. Snowden-Ifft et al., Phys. Rev. D \textbf{61}(2000)101301
\bibitem{DMTPC} S. Ahlen et al., Phys. Lett. B \textbf{695}(2011)124-129
\bibitem{MIMAC} D. Santos et al.,  arxiv:1304.2255(2013)
\bibitem{Nakamura} K. Nakamura et al., PTEP(2015)043F01s
\bibitem{NITPC} C.J. Martoff et al., Nucl. Instr. and Meth. A \textbf{440}(2000)355-359
\bibitem{DRIFT_2} S. Burgos et al., Nucl. Instr. and Meth. A \textbf{600}(2009)417-423
\bibitem{Battat} J.B.R. Battat et al., Physics of the Dark Universe 9-10(2015)1-7
\bibitem{NI} D.R. Nygren J.Phys.Conf.Ser \textbf{65}(2007)012003
\bibitem{nitro} A.R. Prieskorn et al., IEEE Trans.Nucl.Sci. \textbf{58}(2001)2055-2059
\bibitem{Nguyen} Nguyen Phan, Eric Lee in: Cygnus 2015 Conference in Los Angeles
\bibitem{u-PIC} A. Takada et al., Nucl. Instr. Meth. A \textbf{573}(2007)195
\bibitem{GEM} F. Sauli, Sharma, Ann. Rev. Nucl. Part Sci. \textbf{49}(1999)341
\bibitem{SRIM} J.F. Ziegler, J.P. Biersack SRIM The Stopping and Range of Ions in Matter, Code(1985)
\end{thebibliography}
\end{document}